\documentclass[12pt,a4paper]{article}
\usepackage{epsf}
\usepackage{bm}
\makeatletter

\@addtoreset{equation}{section}
\makeatother
\def\a{\alpha}
\def\b{\beta}
\def\ta{\tilde{\alpha}}
\def\tb{\tilde{\beta}}

\def\o{\omega}
\def\th{\theta}
\def\k{\kappa}
\def\t{\tau}
\def\e{\epsilon}
\def\G{\Gamma}
\def\vth{\vartheta}
\newcommand{\Aa}{{\cal A}}
\newcommand{\Bb}{{\cal B}}
\newcommand{\tA}{\tilde{\cal A}}
\newcommand{\tB}{\tilde{\cal B}}
\newcommand{\tc}{\tilde{c}}
\newcommand{\tN}{\tilde{N}}
\newcommand{\tK}{\tilde{K}}

\newcommand{\del}{\partial}

\newcommand{\half}{{1 \over 2}}
\newcommand{\absv}[1]{\left|#1\right|}
\newcommand{\gtsim}{\mathrel{\hbox{\raise0.2ex
\hbox{$>$}\kern-0.75em\raise-0.9ex\hbox{$\sim$}}}}
\newcommand{\ltsim}{\mathrel{\hbox{\raise0.2ex
\hbox{$<$}\kern-0.75em\raise-0.9ex\hbox{$\sim$}}}}
\newcommand{\lw}[1]{\smash{\lower2.0ex\hbox{#1}}}
\newcommand{\commute}[2]{{\left[#1,#2\right]}}

\newcommand{\Tr}{{\rm Tr}}

\newcommand{\bra}[1]{\langle #1 |}
\newcommand{\ket}[1]{| #1 \rangle}

\newcommand{\kslash}{k\kern-0.5em\raise 0.14ex\hbox{/}}
\newcommand{\PRD}[3]{Phys. Rev. {\bf D{#1}} (19{#2}) {#3}}
\newcommand{\PRLet}[3]{Phys. Rev. Lett. {\bf {#1}} (19{#2}) {#3}}

\makeatletter

\@addtoreset{equation}{section}
\makeatother
\begin{document}
\begin{titlepage}
\begin{flushright}
SAGA--HE--168--00\\
October~23,~2000
\end{flushright}
\vspace{50pt}
\centerline{\Large{\bf Charge Generation in the Oscillating Background}}
\vskip2.0cm
\begin{center}
{\bf Koichi~Funakubo$^{a,}$\footnote{e-mail: funakubo@cc.saga-u.ac.jp},
 Akira~Kakuto$^{b,}$\footnote{e-mail: kakuto@fuk.kindai.ac.jp},\\
 Shoichiro~Otsuki$^{b,}$\footnote{e-mail: otk-s1ro@topaz.ocn.ne.jp}
 and Fumihiko~Toyoda$^{b,}$\footnote{e-mail: ftoyoda@fuk.kindai.ac.jp}}
\end{center}
\vskip 1.0 cm
\begin{center}
{\it $^{a)}$Department of Physics, Saga University,
Saga 8408502 Japan}
\vskip 0.2 cm
{\it $^{b)}$Kyushu School of Engineering, Kinki University,
Iizuka 8208555 Japan}
\end{center}
\vskip 1.5 cm
\centerline{\bf Abstract}
\vskip 0.2 cm
\baselineskip=15pt
The preheating after the inflation, which can be interpreted as particle 
creation in the oscillating inflaton background, provides a state far from
thermal equilibrium. We extend the field theoretical treatment of the preheating
by Linde et al. to the case of multicomponent complex scalars to show that
some charges are created in this process, if $C$ and $CP$ are violated.
A new possibility of baryogenesis based on this mechanism is also discussed.
\end{titlepage}
\baselineskip=18pt
\setcounter{page}{2}
\setcounter{footnote}{0}
\section{Introduction}
In the scenario of inflationary cosmology, which can explain the unanswered
problems left by the standard big bang scenario, all the particles at present are
thought to be created in the reheating process following the exponentially
inflating de Sitter phase. That process is viewed as the decay of the inflaton field,
which begins to oscillate around the minimum of the inflaton potential.
The inflaton itself is treated as a coherently oscillating classical field,
and its oscillation is damped by its decay, whose decay products fill the universe to
realize the hot universe.
A several yeas ago, it was pointed out by Linde et al.\cite{KLS1} that
light bosons could be produced more drastically than the perturbative reheating,
when some conditions are satisfied on their coupling to the inflaton and
the amplitudes and frequency of the inflaton oscillation.
This process is called {\it preheating}, to distinguish from the perturbative
reheating.\par
The preheating is nothing but particle creation in the oscillating background.
Although particle creation or annihilation generally occurs in an time-dependent 
background, the number of created particles grows exponentially when the
background is periodic in time. This is a kind of parametric resonance.
This process of particle creation is far from equilibrium, and the state after
that has large quantum fluctuation composed of low frequency modes, whose
relaxation process to the equilibrium is also out of equilibrium.
Hence there are chances for the matter-antimatter asymmetry of the universe
to be generated in these eras, if $C$ and $CP$ are violated during these processes.
In fact, the preheating has been argued in connection to generation of the baryon
asymmetry of the universe (BAU)\cite{Shapsh}, focusing on the fact that
the low-frequency quantum fluctuations of the created particles are so large that
the sphaleron processes may be enhanced.
Besides this possibility, it is natural to think that the baryon number or other
charges were generated during the preheating when all the particles in the universe
were created.
We investigate this possibility to generate a charge asymmetry in the presence of
$C$ and $CP$ violation in the oscillating background.\par
We study the particle creation when $C$ and $CP$ is violated in the interaction
of quantum scalar fields with the background.
We extend the analysis by Linde et al.\cite{KLS2} to a system of multicomponent 
complex scalar fields. The interaction with the oscillating background is approximated
by the successive scattering with some potential, while the effects of $CP$ violation
are evaluated perturbatively.
In \cite{KLS2}, the number density of the created particles during the preheating
was expressed in terms of the wave function in the inflaton background, and
the results were reproduced by the picture of successive scattering approximation.
It is, however, difficult to distinguish the positive-frequency wave from the 
negative-frequency wave in the oscillating background, so that one cannot define
a charge, to which the contributions from the antiparticles have opposite sign
relative to the particles, in terms of wave functions.
Then the treatment based on the successive Bogoliubov transformations will
be useful. In Section~2, we generalize the field theory of a real scalar in
the oscillating background studied in \cite{KLS2} to the case of multicomponent
complex scalars and formulate the descent equations for the Bogoliubov coefficients.
We apply the method to a simple toy model and find some charges to be generated,
in Section 3. The final section is devoted to discussions.
\section{Field Theoretical Treatment}
Consider a system of $n$ complex scalar fields interacting with
the inflaton through the effective lagrangian
\begin{equation}
 {\cal L} =
 \del_\mu\chi_a^*\del^\mu\chi_a -m_a^2(t)\chi_a^*\chi_a
 - \chi_a^* V_{ab}(t) \chi_b
 -\half\left( \chi_a W_{ab}(t) \chi_b +\mbox{c.c.}\right),
                      \label{eq:model-lagrangian}
\end{equation}
where $m_a^2(t)=g_a^2\phi^2(t)$ and $a,b=1,2,\cdots,n$.
The coefficients $V_{ab}(t)$ and $W_{ab}(t)$ represent the collective effects
of the oscillating background and interactions with other fields.
We treat their effects as perturbation.
For simplicity, we consider the case of the undamped background oscillation,
with the parameters in the broad resonance regime.
That is, we assume that the inflaton field is given by
\begin{equation}
 \phi(t) = \Phi\,\sin mt,
\end{equation}
where $m$ is the inflaton mass and $\Phi$ is the constant amplitude, and
neglect the redshift of frequency of each mode.
In this case, each mode of the unperturbed wave functions satisfies
the mode equation
\begin{equation}
 \left(\del_t^2+k^2+g_a^2\Phi^2\sin^2 mt\right)\chi_{ak}(t)=0.
                                         \label{eq:mode-eq-1}
\end{equation}
It is known that wave functions in any periodic potential belong to
either stability or instability bands\cite{LandauQM}.  
The solutions to (\ref{eq:mode-eq-1}) are given by the Mathieu functions,
which have this property.
Among the solutions, those in the instability bands increase their particle numbers
exponentially. 
In particular, for $q_a\equiv{{g_a^2\Phi^2}\over{4m^2}}\gg1$,
some modes are in the broad instability bands so that they continue to grow
for a long time, even if we take the effects of the expanding universe into account.
Further, the frequencies of the modes in the broad resonance band are much larger
than that of the inflaton oscillation. 
Then, the particle number changes only when the inflaton field vanishes, while
the $\chi$-fields evolve adiabatically for time intervals with 
$\absv{\phi(t)}\not=0$.
This observation motivates the study of preheating by approximating the oscillating
background as a series of successive scattering by some simple potential.
In \cite{KLS2}, the inflaton field near its zero was replaced by the quadratic 
potential, and the descent equation for the particle number density in some mode
was derived. The particle number density is given in terms of the wave function, 
which is a solution to (\ref{eq:mode-eq-1}) with {\it any} boundary condition.
As for a charge density, we must distinguish the particle from the antiparticle.
This amounts to impose the boundary condition defining either positive- or
negative-frequency mode. Here we express the densities of particle number and
the charges in terms of the Bogoliubov coefficients which relates the operators
before the inflaton begins to oscillate and those in the adiabatic time intervals.
\subsection{Descent equations for the Bogoliubov coefficients}
We assume that the particles are generated only in the short time interval
near $t_j=\pi j/m_a$, where $j$ is a non-negative integer, and that
the violation of the charge conservation and $CP$ symmetry is effective only
within the short interval when $m_a(t)\simeq0$.
Hence, for $t$ with $t_{j-1}\ll t\ll t_j$, $\chi_{ak}(t)$ satisfies the
unperturbed equation (\ref{eq:mode-eq-1}). Let $f_{ak}^j(t)$ be the properly normalized
positive-frequency solution to (\ref{eq:mode-eq-1}) in the adiabatic interval between $t_{j-1}$ and
$t_j$. Then the $\chi$-field operator is expanded as
\begin{equation}
 \chi_{a}(x) = 
 \int d^3\bm{k}\left( a_{a{\bm k}}^j f_{ak}^j(t)e^{i\bm{kx}} +
                      b_{a{\bm k}}^{j\dagger} f_{ak}^{j*}(t)e^{-i\bm{kx}} \right).
                                \label{eq:mode-exp-chi}
\end{equation}
Without the perturbation, upon scattered by the potential $g_a^2\phi^2(t)$ at $t_j$,
the positive-frequency wave $f_{ak}^j(t)$ is transmitted to a linear combination of
$f_{ak}^{j+1}(t)$ and $f_{ak}^{j+1*}(t)$, whose coefficients are determined by
solving the scattering problem by the potential at $t_j$.
In the presence of the perturbation $V_{ab}$ and $W_{ab}$, the scattered wave
is composed of various components of $f_{ak}^{j+1}(t)$ and $f_{ak}^{j+1*}(t)$.
Suppose that the positive-frequency wave $f_{ak}^0(t)$ at $t<t_0$ transfers
to a set of the positive- and negative-frequency waves in the $j$-th interval;
\begin{equation}
 \left(\begin{array}{c} 0 \\ \vdots \\ f_{ak}^0(t) \\ \vdots \\ 0\end{array}
 \right) \longrightarrow \cdots \longrightarrow
 \left(\begin{array}{c}
  \a_{a1}^jf_{1k}^j(t) + \b_{a1}^j f_{1k}^{j*}(t) \\ \vdots\\
  \a_{ab}^jf_{bk}^j(t) + \b_{ab}^j f_{bk}^{j*}(t) \\ \vdots\\
  \a_{an}^jf_{nk}^j(t) + \b_{an}^j f_{nk}^{j*}(t) \end{array}\right),
                             \label{eq:def-alpha-beta-1}
\end{equation}
while the negative-frequency wave $f_{ak}^{0*}(t)$ transfers as
\begin{equation}
 \left(\begin{array}{c} 0 \\ \vdots \\ f_{ak}^{0*}(t) \\ \vdots \\ 0\end{array}
 \right) \longrightarrow \cdots \longrightarrow
 \left(\begin{array}{c}
  \tb_{a1}^jf_{1k}^j(t) + \ta_{a1}^j f_{1k}^{j*}(t) \\ \vdots\\
  \tb_{ab}^jf_{bk}^j(t) + \ta_{ab}^j f_{bk}^{j*}(t) \\ \vdots\\
  \tb_{an}^jf_{nk}^j(t) + \ta_{an}^j f_{nk}^{j*}(t) \end{array}\right),
                             \label{eq:def-alpha-beta-2}
\end{equation}
where $\a_{ab}^j\not=\ta_{ab}^j$ and $\b_{ab}^j\not=\tb_{ab}^j$ because of
the $CP$ violation. Then the Bogoliubov transformation between the creation and
annihilation operators at $t<t_0$ and those at $t_{j-1}<t<t_j$ is
\begin{eqnarray}
 a_{a{\bm k}}^j &=& 
 a_{b{\bm k}}^0\a_{ba}^j + b_{b{\bm k}}^{0\dagger}\tb_{ba}^j, \nonumber\\
 b_{a{\bm k}}^{j\dagger} &=&
 a_{b{\bm k}}^0\b_{ba}^j + b_{b{\bm k}}^{0\dagger}\ta_{ba}^j.
                          \label{eq:op-B-trf-0-j}
\end{eqnarray}
Requiring 
\begin{equation}
 \commute{a_{a{\bm k}}^j}{a_{b{\bm k}'}^{j\dagger}} =
 \commute{b_{a{\bm k}}^j}{b_{b{\bm k}'}^{j\dagger}} = \delta_{ab}\delta^3({\bm k}-{\bm k}'),\qquad
 \commute{a_{a{\bm k}}^j}{b_{b{\bm k}'}^j} = 0,
\end{equation}
we find the conditions on the Bogoliubov coefficients:
\begin{eqnarray}
 &&\a_{ca}^j\a_{cb}^{j*} - \tb_{ca}^j\tb_{cb}^{j*} = \delta_{ab}, \nonumber\\
 &&\ta_{ca}^{j*}\ta_{cb}^j - \b_{ca}^{j*}\b_{cb}^j = \delta_{ab},  \nonumber\\
 &&\a_{ca}^j\b_{cb}^{j*}-\tb_{ca}^j\ta_{cb}^{j*} = 0.
\end{eqnarray}
In terms of the $n$-by-$n$ matrix notation, $\a^j=\left(\a_{ab}^j\right)$,
$\b^j=\left(\b_{ab}^j\right)$, etc., these conditions are expressed as
\begin{equation}
 \a^{j\dagger}\a^j-\tb^{j\dagger}\tb^j =
 \ta^{j\dagger}\ta^j-\b^{j\dagger}\b^j = 1,\qquad
 \b^{j\dagger}\a^j - \ta^{j\dagger}\tb_j = 0.   \label{eq:cond-B-coeff-0}
\end{equation}
We can also show that
\begin{equation}
 \a^j\a^{j\dagger}-\b^j\b^{j\dagger} =
 \ta^j\ta^{j\dagger}-\tb^j\tb^{j\dagger} = 1,\qquad
 \a^j\tb^{j\dagger} - \b^j\ta^{j\dagger} = 0.  \label{eq:cond-B-coeff-2}
\end{equation}
\par
Now we define the number density of the particles with momentum $\bm{k}$ 
in the $j$-th interval as
\begin{equation}
 n_k^j \equiv
 {1\over V}\bra{0^0}\sum_{a=1}^n\left(
  a_{a{\bm k}}^{j\dagger}a_{a{\bm k}}^j +  b_{a{\bm k}}^{j\dagger}b_{a{\bm k}}^j \right)
 \ket{0^0},             \label{eq:def-n-k-j}
\end{equation}
where $\ket{0^0}$ is defined by $a_{a{\bm k}}^0\ket{0^0}=b_{a{\bm k}}^0\ket{0^0}=0$
and $V=\delta^3({\bm k}={\bm0})$.
By use of (\ref{eq:op-B-trf-0-j}), we have
\begin{equation}
 n_k^j = 
 \sum_{a,b}\left(\tb_{ba}^{j*}\tb_{ba}^j+\b_{ba}^j\b_{ba}^{j*}\right) =
 \Tr\left(\tb^{j\dagger}\tb^j + \b^{j\dagger}\b^j\right).
                         \label{eq:n-k-j}
\end{equation}
If each component of $\phi_a(x)$ has the $U(1)$ charge $Q_a$, the density of
the generated charge in mode $\bm{k}$ in the $j$-th interval is given by
\begin{equation}
 j^j_k\equiv
 {1\over V}\bra{0^0}\sum_{a=1}^n Q_a\left(
  a_{a{\bm k}}^{j\dagger}a_{a{\bm k}}^j -  b_{a{\bm k}}^{j\dagger}b_{a{\bm k}}^j \right)
 \ket{0^0},             \label{eq:def-j-k-j}
\end{equation}
which is written in terms of the Bogoliubov coefficients as
\begin{equation}
 j^j_k =
 \sum_{a,b} 
 Q_a\left(\tb_{ba}^{j*}\tb_{ba}^j-\b_{ba}^j\b_{ba}^{j*}\right) =
 \Tr\left[Q\left(\tb^{j\dagger}\tb^j - \b^{j\dagger}\b^j\right)\right],
                          \label{eq:j-k-j}
\end{equation}
where $Q$ in the last line is $n$-by-$n$ matrix,
$Q={\rm diag}(Q_1,Q_2,\cdots,Q_n)$.\par
In the case of $n=1$, it is easy to find $\absv{\b^j}^2=\absv{\tb^j}^2$ 
by use of (\ref{eq:cond-B-coeff-0}), so that $j_k^j=0$.
Hence, we need at least two complex scalars interacting with the oscillating background
in order to generate charge asymmetry.
This fact can be understood as follows: If we regard the particles and the 
charges as the decay products of the oscillating field, at least two decay channels
are needed to produce nonzero charge since the effect of $CP$ violation appears
as the interference of the two channels. This is the same situation as the 
GUT-baryogenesis by the heavy boson decay.\par
The Bogoliubov coefficients in the $(j+1)$-th interval are related to those in the
$j$-th interval by solving the scattering problem by the potential at $t_j$, 
just as done in \cite{KLS2}.
Within each adiabatic interval, the positive frequency wave function $f_{ak}^j(t)$
is well approximated by
\begin{equation}
 h_{ak}(t) = {1\over\sqrt{2\omega_a(t)}}\,e^{-i\int_0^t dt'\,\omega_a(t')},
                     \label{eq:def-adiabatic-wave}
\end{equation}
where
\begin{equation}
 \omega_a(t) = \sqrt{k^2+m_a^2(t)}=\sqrt{k^2 + g_a^2\Phi^2\sin^2mt},  
                                    \label{eq:def-omega-t}
\end{equation}
up to some constant phase.
On the other hand, for $t$ near $t_j$, for which the adiabatic approximation is 
no longer valid, we solve the scattering problem employing some approximate form
of the potential at $t_j$.
We adopt another type of potential which allows to define an asymptotic state, that is, 
$\omega_a(t)\simeq \omega_a^0=\mbox{const.}$ for $\absv{t-t_j}\gg1/m$, in contrast
to the quadratic potential $m_a^2(t)\simeq m^2(t-t_j)^2$ used in \cite{KLS2}.
This choice of the potential is more appropriate for the perturbative calculation
of the $CP$-violating scattering presented in the next subsection.
The general formulas presented in this subsection holds independent of the detailed
form of the potential.
If we use this potential, the adiabatic wave (\ref{eq:def-adiabatic-wave}) behaves, 
far from the potential, as
\begin{equation}
 h_{ak}(t)\;\rightarrow\;
 {1\over\sqrt{2\o_a^0}}\,e^{-i\o_a^0(t-t_j)-i\th_{ak}^j-i\vth_{ak}^\pm},
 \qquad (\mbox{as }t-t_j\rightarrow \pm\infty) \label{eq:asymp-h}
\end{equation}
where $\th_{ak}^j=\int_0^{t_j}dt'\,\o_a(t')$ is the phase accumulated by the
moment $t_j$ and $\vth_{ak}^+$ ($\vth_{ak}^-$) is the phase difference between 
$\int_t^{t_j}dt'\,\o_a(t')$ and $\o_a^0(t-t_j)$ for $t\gg t_j$ ($t\ll t_j$).
Then the positive-frequency wave at $t<t_0$, $f_{ak}^0(t)$, can be approximated by
$h_{ak}(t)$, while $f_{ak}^j(t)$ in the $j$-th adiabatic interval
will be expressed as a linear combination of $h_{bk}(t)$ and $h_{bk}^*(t)$.\par
Suppose that we find the solutions to the wave equations (usually,
by perturbative method) satisfying the boundary condition that
only the positive-frequency wave in the $a$-th component exists at $t\ll t_j$;
\begin{equation}
 \pmatrix{0\cr\vdots\cr{1\over\sqrt{2\o_a^0}}\,e^{-i\o_a^0(t-t_j)}\cr \vdots \cr 0}
 \stackrel{t\ll t_j}{\longleftarrow}
 \left(\begin{array}{c}
   \chi_{1k}(t)\\ \vdots \\ \chi_{ak}(t) \\ \vdots \\ \chi_{nk}(t)
       \end{array}\right)
 \stackrel{t\gg t_j}{\longrightarrow}
 \pmatrix{
  \Aa_{a1}{1\over\sqrt{2\o_1^0}}\,e^{-i\o_1^0(t-t_j)} +
  \Bb_{a1}{1\over\sqrt{2\o_1^0}}\,e^{i\o_1^0(t-t_j)} \cr \vdots \cr
  \Aa_{ab}{1\over\sqrt{2\o_b^0}}\,e^{-i\o_b^0(t-t_j)} +
  \Bb_{ab}{1\over\sqrt{2\o_b^0}}\,e^{i\o_b^0(t-t_j)} \cr \vdots \cr
  \Aa_{an}{1\over\sqrt{2\o_n^0}}\,e^{-i\o_n^0(t-t_j)} +
  \Bb_{an}{1\over\sqrt{2\o_n^0}}\,e^{i\o_n^0(t-t_j)} },
                             \label{eq:scattered-pos-freq}
\end{equation}
and similarly the solution satisfying the boundary condition that
only the negative-frequency wave exists at $t\ll t_j$;
\begin{equation}
 \pmatrix{0\cr\vdots\cr{1\over\sqrt{2\o_a^0}}\,e^{i\o_a^0(t-t_j)}\cr \vdots \cr 0}
 \stackrel{t\ll t_j}{\longleftarrow}
 \left(\begin{array}{c}
   \chi_{1k}(t)\\ \vdots \\ \chi_{ak}(t) \\ \vdots \\ \chi_{nk}(t)
       \end{array}\right)
 \stackrel{t\gg t_j}{\longrightarrow}
 \pmatrix{
  \tB_{a1}{1\over\sqrt{2\o_1^0}}\,e^{-i\o_1^0(t-t_j)} +
  \tA_{a1}{1\over\sqrt{2\o_1^0}}\,e^{i\o_1^0(t-t_j)} \cr \vdots \cr
  \tB_{ab}{1\over\sqrt{2\o_b^0}}\,e^{-i\o_b^0(t-t_j)} +
  \tA_{ab}{1\over\sqrt{2\o_b^0}}\,e^{i\o_b^0(t-t_j)} \cr \vdots \cr
  \tB_{an}{1\over\sqrt{2\o_n^0}}\,e^{-i\o_n^0(t-t_j)} +
  \tA_{an}{1\over\sqrt{2\o_n^0}}\,e^{i\o_n^0(t-t_j)} }.
                             \label{eq:scattered-neg-freq}
\end{equation}
Here the coefficients $\Aa$, $\Bb$, $\tA$ and $\tB$, which are also kinds of
Bogoliubov coefficients, satisfy the same relations as $\a$, $\b$, $\ta$ and $\tb$:
\begin{eqnarray}
 &&\Aa^\dagger\Aa-\tB^\dagger\tB=\tA^\dagger\tA-\Bb^\dagger\Bb=1,\qquad
   \Bb^\dagger\Aa-\tA^\dagger\tB = 0,   \nonumber\\
 &&\Aa\Aa^\dagger-\Bb\Bb^\dagger=\tA\tA^\dagger-\tB\tB^\dagger=1,\qquad
   \Aa\tB^\dagger-\Bb\tA^\dagger=0.  \label{eq:conditions-A-B}
\end{eqnarray}
Inverting (\ref{eq:def-alpha-beta-1}) and (\ref{eq:def-alpha-beta-2}) and
replacing $f_{ak}^0(t)$ with $h_{ak}(t)$, we have
\begin{equation}
 \pmatrix{0 \cr \vdots \cr f_{ak}^j(t) \cr \vdots \cr 0}
 \sim
 \pmatrix{
  \left(\a^{j\dagger}\right)_{a1} h_{1k}(t) -
  \left(\tb^{j\dagger}\right)_{a1} h_{1k}^*(t) \cr \vdots \cr
  \left(\a^{j\dagger}\right)_{ab} h_{bk}(t) -
  \left(\tb^{j\dagger}\right)_{ab} h_{bk}^*(t) \cr \vdots \cr
  \left(\a^{j\dagger}\right)_{an} h_{nk}(t) -
  \left(\tb^{j\dagger}\right)_{an} h_{nk}^*(t) },
         \label{eq:fj-to-0}
\end{equation}
and
\begin{equation}
 \pmatrix{0 \cr \vdots \cr g_{ak}^{j*}(t) \cr \vdots \cr 0}
 \sim
 \pmatrix{
  \left(\ta^{j\dagger}\right)_{a1} h_{1k}^*(t) -
  \left(\b^{j\dagger}\right)_{a1} h_{1k}(t) \cr \vdots \cr
  \left(\ta^{j\dagger}\right)_{ab} h_{bk}^*(t) -
  \left(\b^{j\dagger}\right)_{ab} h_{bk}(t) \cr \vdots \cr
  \left(\ta^{j\dagger}\right)_{an} h_{nk}^*(t) -
  \left(\b^{j\dagger}\right)_{an} h_{nk}(t) }.
         \label{eq:gj-to-0}
\end{equation}
Since how $h_{ak}(t)$ and $h_{ak}^*(t)$ appearing in each element in the \
right-hand sides of (\ref{eq:fj-to-0}) and (\ref{eq:gj-to-0}) are scattered by 
the potential at $t_j$ are determined by
(\ref{eq:scattered-pos-freq}) and (\ref{eq:scattered-neg-freq}), respectively,
the $b$-th component of the right-hand side of (\ref{eq:fj-to-0}) becomes,
after the scattering ($t\gg t_j$),
\begin{eqnarray}
 &&
 \left[
  \left(\a^{j\dagger}\right)_{ac}\,e^{-i(\th_{ck}^j+\vth_{ck}^-)}\Aa_{cb}
 -\left(\tb^{j\dagger}\right)_{ac}\,e^{i(\th_{ck}^j+\vth_{ck}^-)}\tB_{cb}
 \right]\,e^{i(\th_{bk}^j+\vth_{bk}^+)}\, h_{bk}(t)  \nonumber\\
 &&\qquad-
 \left[
  \left(\tb^{j\dagger}\right)_{ac}\,e^{i(\th_{ck}^j+\vth_{ck}^-)}\tA_{cb}
 -\left(\a^{j\dagger}\right)_{ac}\,e^{-i(\th_{ck}^j+\vth_{ck}^-)}\Bb_{cb}
 \right]\,e^{-i(\th_{bk}^j+\vth_{bk}^+)}\, h_{bk}^*(t)
                        \label{eq:scattered-f-j-a}
\end{eqnarray}
which should be identified with
$
  \left(\a^{j+1\dagger}\right)_{ab} h_{bk}(t) -
  \left(\tb^{j+1\dagger}\right)_{ab} h_{bk}^*(t),
$
and the $b$-th component of the right-hand-side of (\ref{eq:gj-to-0}) becomes
\begin{eqnarray}
 &&
 \left[
  \left(\ta^{j\dagger}\right)_{ac}\,e^{i(\th_{ck}^j+\vth_{ck}^-)}\tB_{cb}
 -\left(\b^{j\dagger}\right)_{ac}\,e^{-i(\th_{ck}^j+\vth_{ck}^-)}\Aa_{cb}
 \right]\,e^{i(\th_{bk}^j+\vth_{bk}^+)}\, h_{bk}(t)  \nonumber\\
 &&\qquad-
 \left[
  \left(\b^{j\dagger}\right)_{ac}\,e^{-i(\th_{ck}^j+\vth_{ck}^-)}\Bb_{cb}
 -\left(\ta^{j\dagger}\right)_{ac}\,e^{i(\th_{ck}^j+\vth_{ck}^-)}\tA_{cb}
 \right]\,e^{-i(\th_{bk}^j+\vth_{bk}^+)}\, h_{bk}^*(t).
                        \label{eq:scattered-g-j-a}
\end{eqnarray}
which should be identified with
$
 \left(\ta^{j+1\dagger}\right)_{ab} h_{bk}^*(t) -
  \left(\b^{j+1\dagger}\right)_{ab} h_{bk}(t).
$
If we denote the $n$-by-$n$ matrix of the phase factors as
\begin{equation}
 U_{\pm}= \mbox{diag}\left(
  e^{-i(\th_{1k}^j+\vth_{1k}^\pm)},\cdots,\,e^{-i(\th_{nk}^j+\vth_{nk}^\pm)}
          \right),
\end{equation}
we obtain the descent equations for the Bogoliubov coefficients
\begin{eqnarray}
 \a^{j+1} &=&
 U_+\left(\Aa^\dagger U_-^\dagger\a^j-\tB^\dagger U_- \tb^j\right)
                                              \nonumber\\
 \b^{j+1} &=&
 U_+\left(\Aa^\dagger U_-^\dagger\b^j-\tB^\dagger U_-\ta^j \right)
                                              \nonumber\\
 \ta^{j+1} &=&
 U_+^\dagger\left(\tA^\dagger U_-\ta^j-\Bb^\dagger U_-^\dagger\b^j\right)
                                              \nonumber\\
 \tb^{j+1} &=&
 U_+^\dagger\left(\tA^\dagger U_-\tb^j-\Bb^\dagger U_-^\dagger\a^j\right).
                               \label{eq:ab-j+1-by-j}
\end{eqnarray}
As a check, one can show that these $(j+1)$-th Bogoliubov coefficients satisfies 
the same relations as (\ref{eq:cond-B-coeff-0}) and (\ref{eq:cond-B-coeff-2}).\par
\subsection{Perturbative calculation of the scattering matrices}
Now we calculate the scattering matrices $\Aa$, $\Bb$, $\tA$ and $\tB$ by
the perturbative method.
The wave equations to be solved are written as
\begin{equation}
 \left[
  \left(\del_t^2+k^2+m_a^2(t)\right){\bf 1} +
  \pmatrix{ V_{ab}(t) & W_{ab}^*(t) \cr
            W_{ab}(t) & V_{ab}^*(t) }
 \right]\pmatrix{\chi_{bk}(t)\cr \chi_{bk}^*(t)} = 0,
\end{equation}
For convenience, let us introduce the following symbols:
\begin{eqnarray}
 X_k(t) &=& \pmatrix{\chi_{ak}(t)\cr \chi_{ak}^*(t)}, \nonumber\\
 \Delta_0 &=& \left(\del_t^2+k^2+m_a^2(t)\right){\bf 1}=
 \left(\del_t^2+\omega_a^2(t) \right){\bf 1},  \\
 \Delta_1 &=&
 \pmatrix{ V_{ab}(t) & W_{ab}^*(t) \cr
            W_{ab}(t) & V_{ab}^*(t) }.   \nonumber
\end{eqnarray}
Then a perturbative solution to the equation
\begin{equation}
 \left(\Delta_0+\Delta_1\right)X_k(t)=0,
\end{equation}
to the first order with respect to $\Delta_1$ is given by
\begin{equation}
 X_k(t) = X_k^{(0)}(t) - \Delta_0^{-1}\Delta_1X_k^{(0)}(t),
                         \label{eq:perturb-sol}
\end{equation}
where $X_k^{(0)}(t)$ is a solution to
\begin{equation}
 \Delta_0 X_k^{(0)}(t)=0.  \label{eq:unperturb-eq}
\end{equation}
In order to evaluate the scattering matrices $\Aa$ and $\Bb$, we must prepare
$X_k(t)$ such that only $\chi_{ak}(t)$ is the positive-frequency wave and
the other components vanish in the remote past.
We solve the scattering problem with this initial condition for the approximate
potential which admits the asymptotic states.
Let $f_{ak}(t)$ be the wave function satisfying
\begin{equation}
 \left(\del_t^2+\omega_a^2(t)\right)f_{ak}(t) = 0,\quad\mbox{and}\quad
 f_{ak}(t)\stackrel{t\rightarrow-\infty}{\longrightarrow}
 {1\over\sqrt{2\omega_a^0}}e^{-i\omega_a^0 t}.
                             \label{eq:def-f-ak}
\end{equation}
Suppose that we solve this unperturbed equation and find the expression as
\begin{equation}
 f_{ak}(t) = \a_af_{ak}^{\rm out}(t) + \b_af_{ak}^{\rm out*}(t),
                     \label{eq:rel-f-f-out}
\end{equation}
where $f_{ak}^{\rm out}(t)$ satisfies
\begin{equation}
 \left(\del_t^2+\omega_a^2(t)\right)f_{ak}^{\rm out}(t) = 0,
 \quad\mbox{and}\quad
 f_{ak}^{\rm out}(t)\stackrel{t\rightarrow\infty}{\longrightarrow}
 {1\over\sqrt{2\omega_a^0}}e^{-i\omega_a^0 t}.
                             \label{eq:def-f-ak-out} 
\end{equation}
According to the general theory of distorted-wave Born approximation\cite{FKOTT},
the Green function $G_a(t,t')$ satisfying
\begin{equation}
 \left(\del_t^2+\omega_a^2(t)\right) G_a(t,t') = \delta(t-t'),
\end{equation} 
with the boundary condition such that there is only positive-frequency
wave in the remote past is given by
\begin{equation}
 G_a(t,t') = \left\{
  \begin{array}{ll}
   {1\over w}\,f_{ak}(t)f_{ak}^*(t'), & (t<t') \\
   {1\over w}\,f_{ak}(t')f_{ak}^*(t), & (t'<t)
  \end{array}\right.
\end{equation}
where $w$ is the Wronskian defined by
\begin{equation}
 w\equiv f_{ak}(t)\del_t f_{ak}^*(t) - f_{ak}^*(t)\del_t f_{ak}(t),
\end{equation}
which is time-independent and $w=i$ in our case. 
Then our $\Delta_0^{-1}$ is
\begin{equation}
 \Delta_0^{-1} =
 \mbox{diag}\left( G_1, \cdots, G_n, G_1^*,\cdots, G_n^*\right).
\end{equation}
\par
As the unperturbed wave $X_k^{(0)}(t)$, we take
\begin{equation}
 X_k^{[a](0)}(t) =
 \pmatrix{ cf_k(t) \cr c^*f_k^*(t)} =
 \pmatrix{ c_{a1}f_{1k}(t)\cr\vdots \cr c_{a1}^*f_{1k}^*(t)\cr\vdots},
\end{equation}
and chose the set of parameters $c_{ab}$ such that
\begin{equation}
 X_k^{[a]}(t) =
 X_k^{[a](0)}(t) - \Delta_0^{-1}\Delta_1 X_k^{[a](0)}(t)
 \stackrel{t\rightarrow-\infty}{\longrightarrow}
 \pmatrix{ 0\cr\vdots\cr f_{ak}(t)\cr\vdots\cr f_{ak}^*(t)\cr\vdots\cr 0 }.
                 \label{eq:IC-on-X-a}
\end{equation}
In terms of the $n$-component wave function, the perturbed wave becomes
\begin{equation}
 \chi_k^{[a]}(t) =
 cf_k(t) - GV(cf_k)(t) - GW^*(c^*f_k^*)(t).  \label{eq:perturb-chi-a}
\end{equation}
Requiring the initial condition (\ref{eq:IC-on-X-a}), we find that $c_{ab}$
should be taken to satisfy
\begin{equation}
 c_{ab} +
 i c_{ac}\int_{-\infty}^\infty dt'\,V_{bc}(t')
      f_{bk}^*(t')f_{ck}(t') +
 i c_{ac}^*\int_{-\infty}^\infty dt'\,W_{bc}^*(t')
      f_{bk}^*(t') f_{ck}^*(t')
 =
 \delta_{ab}.                   \label{eq:condition-c-ab}
\end{equation}
Once $c_{ab}$ is determined, we obtain the scattering matrices $\Aa$ and $\Bb$
by extracting the coefficients of the positive- and negative-frequency waves,
respectively, from (\ref{eq:perturb-chi-a}) in the remote future:
\begin{eqnarray}
 \Aa_{ab} &=&
 c_{ab}\a_b + i \b_b^*
 \int_{-\infty}^\infty dt'\,f_{bk}(t')\left[
  V_{bc}(t')c_{ac}f_{ck}(t')+W_{bc}^*(t')c_{ac}^*f_{ck}^*(t')\right],
                                               \nonumber\\
 \Bb_{ab} &=&
 c_{ab}\b_b + i\a_b^*
 \int_{-\infty}^\infty dt'\,f_{bk}(t')\left[
  V_{bc}(t')c_{ac}f_{ck}(t')+W_{bc}^*(t')c_{ac}^*f_{ck}^*(t')\right].
                                  \label{eq:general-A-B}
\end{eqnarray}
Similarly, by preparing the negative-frequency wave in the remote past,
we can find the other set of the scattering matrices $\tA$ and $\tB$.
They are given by
\begin{eqnarray}
 \tA_{ab} &=&
 \tc_{ab}\a_b^* - i \b_b
 \int_{-\infty}^\infty dt'\,f_{bk}^*(t')\left[
  V_{bc}(t')\tc_{ac}f_{ck}^*(t')+W_{bc}^*(t')\tc_{ac}^*f_{ck}(t')\right],
                                               \nonumber\\
 \tB_{ab} &=&
 \tc_{ab}\b_b^* - i\a_b 
 \int_{-\infty}^\infty dt'\,f_{bk}^*(t')\left[
  V_{bc}(t')\tc_{ac}f_{ck}^*(t')+W_{bc}^*(t')\tc_{ac}^*f_{ck}(t')\right],
                                  \label{eq:general-tA-tB}
\end{eqnarray}
where $\tc_{ab}$ must satisfy
\begin{equation}
 \tc_{ab} -
 i \tc_{ac}\int_{-\infty}^\infty dt'\,V_{bc}(t')
               f_{bk}(t') f_{ck}^*(t') -
 i \tc_{ac}^*\int_{-\infty}^\infty dt'\,W_{bc}^*(t')
               f_{bk}(t') f_{ck}(t')
 = \delta_{ab}.             \label{eq:condition-tc-ab}
\end{equation}
\par
In the case of $W_{ab}(t)=0$, for which the global $U(1)$ symmetry exists
corresponding to the rotation of all $\chi_a(x)$ with a common phase,
the scattering matrices are simply given by
\begin{eqnarray}
 \Aa &=& N^{-1}\left(\a+K\b^\dagger\right),\qquad
 \Bb  =  N^{-1}\left(\b+K\a^\dagger\right),  \nonumber\\
 \tA &=& \tN^{-1}\left(\a^\dagger+\tK\b\right),\qquad
 \tB  =  \tN^{-1}\left(\b^\dagger+\tK\a\right),
                      \label{eq:B-matrix-for-zero-W}
\end{eqnarray}
where $\a$ and $\b$ are diagonal matrices defined by
\begin{equation}
 \a = \mbox{diag}\left(\a_1,\a_2,\cdots,\a_n\right),\qquad
 \b = \mbox{diag}\left(\b_1,\b_2,\cdots,\b_n\right),
\end{equation}
and the matrices $N$, $\tN$, $K$ and $\tK$ are given by
\begin{eqnarray}
 N_{ab} &=&
 \delta_{ab} + i\int_{-\infty}^\infty dt\,
             V_{ba}(t)f_{ak}(t)f_{bk}^*(t),          \nonumber\\
 \tN_{ab} &=&
 \delta_{ab} - i\int_{-\infty}^\infty dt\,
             V_{ba}(t)f_{ak}^*(t)f_{bk}(t),    \label{eq:def-N-tN}\\ 
 K_{ab} &=&
 i\int_{-\infty}^\infty dt\,V_{ba}(t)f_{ak}(t)f_{bk}(t),
                                           \nonumber\\
 \tK_{ab} &=&
 -i\int_{-\infty}^\infty dt\,V_{ba}(t)f_{ak}^*(t)f_{bk}^*(t).
                                       \label{eq:def-K-tK}
\end{eqnarray}
Because of the hermiticity of the matrix $V_{ab}(t)$, $\tK^\dagger=K$ and
$T_{ab}\equiv\int_{-\infty}^\infty dt\,V_{ba}(t)f_{ak}(t)f_{bk}^*(t)=T_{ba}^*$,
the matrix $N$ can be regarded as the unitary matrix $\exp(iT)$.

\section{Example}
Once the model of the inflaton interacting with at least two complex scalars is
specified, one can evaluate the `effective potential' $V_{ab}(t)$ and $W_{ab}(t)$,
which may be induced directly by coupling with the inflaton to the complex scalars,
or through some loop effects including the scalars and any other fields.
Here, without proposing any specific model, we illustrate calculation of the 
generated charges and particle numbers, assuming some simple form of the 
$CP$-violating potentials.
For simplicity, we concentrate on the case of $W_{ab}(t)=0$ and $n=2$ and
assume the degeneracy of the scalar masses ($m_1=m_2$) and
the property of the potential $V_{11}=V_{22}$, which may result from 
some discrete symmetry. Under these assumptions, 
\begin{equation}
 f_{1k}(t) = f_{2k}(t) = f_k(t),\qquad
 \a_1=\a_2=\a,\quad \b_1=\b_2=\b.
\end{equation}
Then the matrices in (\ref{eq:def-N-tN}) and (\ref{eq:def-K-tK}) are reduced to
\begin{eqnarray}
 &&K =\pmatrix{iK_1 & iK_2(1-i\epsilon_K) \cr
               iK_2(1+i\epsilon_K) & iK_1 },\qquad
   \tK = K^\dagger,    \nonumber\\
 &&N=\pmatrix{ 1+iN_1 & iN_2(1-i\epsilon_N) \cr
               iN_2(1+i\epsilon_N) & 1+iN_1 },\qquad
 \tN= N^\dagger,     \label{eq:2-dim-K-N}
\end{eqnarray}
where
\begin{eqnarray}
 &&
 N_1=\int_{-\infty}^\infty dt\,V_{11}(t)\absv{f_k(t)}^2,\qquad
 N_2=\int_{-\infty}^\infty dt\,{\rm Re}V_{12}(t)\absv{f_k(t)}^2,\nonumber\\
 &&
 \epsilon_N =
 {1\over{N_2}}\int_{-\infty}^\infty dt\,{\rm Im}V_{12}(t)\absv{f_k(t)}^2,
                                             \label{eq:def-epsilon-N}\\
 &&
 K_1=\int_{-\infty}^\infty dt\,V_{11}(t) f_k^2(t), \qquad
 K_2=\int_{-\infty}^\infty dt\,{\rm Re}V_{12}(t)\,f_k^2(t),     \nonumber\\
 &&
 \epsilon_K =
 {1\over{K_2}}\int_{-\infty}^\infty dt\,{\rm Im}V_{12}(t)\,f_k^2(t).
                                             \label{eq:def-epsilon-K}
\end{eqnarray}
Once the explicit forms of $f_k(t)$, $V_{11}(t)$ and $V_{12}(t)$ are provided,
we can calculate all the quantities needed to follow the descent equations
(\ref{eq:ab-j+1-by-j}), so that the time evolution of the densities of
particle number and charges.\par
First of all, we prepare the unperturbed wave which is appropriate for
the perturbative calculation of the scattering matrices.
For this purpose, we substitute the oscillating potential near its zeros with 
the asymptotically flat potential as
\begin{equation}
 \Phi^2\sin^2 mt \simeq 
 2\Phi^2\tanh^2\left({{m(t-t_j)}\over\sqrt{2}}\right). \label{eq:approx-potential}
\end{equation}
This is better than the simple quadratic potential adopted in \cite{KLS2},
in that the potential of (\ref{eq:approx-potential}) matches to the original
oscillating potential up to the forth order of $m(t-t_j)$.
Introducing the dimensionless parameters
\begin{equation}
 \t = k_*(t-t_j),\quad \k = {k\over{k_*}},\quad
 \xi = \sqrt{{m\over{2g\Phi}}} = {1\over{2q^{1/4}}},
\end{equation}
where $k_*=\sqrt{g\Phi m}=\sqrt{2}mq^{1/4}$ and $q={{g^2\Phi^2}\over{4m^2}}$,
the wave equation is expressed as
\begin{equation}
 {{d^2f_k(\t)}\over{d\t^2}} +
 \left[\k^2+{1\over{\xi^2}}\tanh^2(\xi\t)\right] f_k(\t) = 0.
                              \label{eq:wave-eq-fk}
\end{equation}
We find that the solution which asymptotically approaches the positive-frequency 
wave in the remote past is given, in terms of $x\equiv \xi\t$, by
\begin{eqnarray}
 f_k(t) &=&
 {1\over\sqrt{2\omega_0}}\left(e^x+e^{-x}\right)^{-i\e}\,
 F\!\left(a, b, c; {1\over{1+e^{-2x}}} \right)   \nonumber\\
 &&
 \stackrel{\t\rightarrow-\infty}{\longrightarrow}
 {1\over\sqrt{2\omega_0}}e^{-i\omega_0(t-t_j)},
\end{eqnarray}
where
\begin{eqnarray}
 \e &=& {{g\Phi}\over m}\sqrt{1+{{k^2}\over{2g^2\Phi^2}}} =
 2q^{1/2}\sqrt{1+{{\k^2}\over{4q^{1/2}}}},   \nonumber\\
 \omega_0 &=& 2\e\xi k_* = \sqrt2\,m \e,   \\
 a &=& 
 -4iq^{1/2}\left(\sqrt{1+{{\k^2}\over{4q^{1/2}}}}-\sqrt{1-{1\over{64q}}}\right)
  + \half,   \nonumber\\
 b &=&
 -4iq^{1/2}\left(\sqrt{1+{{\k^2}\over{4q^{1/2}}}}+\sqrt{1-{1\over{64q}}}\right)
  + \half,   \nonumber\\
 c &=&
 -4iq^{1/2}\sqrt{1+{{\k^2}\over{4q^{1/2}}}} +1,
\end{eqnarray}
and $F(a,b,c;z)$ is the hypergeometric function.
By use of its property, we find, in the limit of $t-t_j\rightarrow\infty$,
\begin{equation}
 f_k(t) \longrightarrow
 {{\a_k}\over\sqrt{2\omega_0}}e^{-i\omega_0(t-t_j)} +
 {{\b_k}\over\sqrt{2\omega_0}}e^{i\omega_0(t-t_j)},
\end{equation}
where
\begin{eqnarray}
 \a_k &=&
 {{\G(-2i\e+1)\G(-2i\e)}\over
  {\G\left(\half-2i\e+4iq^{1/2}\sqrt{1-{1\over{64q}}}\right)
   \G\left(\half-2i\e-4iq^{1/2}\sqrt{1-{1\over{64q}}}\right)}}, \nonumber\\
 \b_k &=&
 {{\G(-2i\e+1)\G(2i\e)}\over
  {\G\left(\half+4iq^{1/2}\sqrt{1-{1\over{64q}}}\right)
   \G\left(\half-4iq^{1/2}\sqrt{1-{1\over{64q}}}\right)}} =
 -i{{\cosh\left(4\pi q^{1/2}\sqrt{1-{1\over{64q}}}\right)}\over
    {\sinh(2\pi\e)}}. \nonumber\\
                                   \label{eq:unperturb-a-b}
\end{eqnarray}
Among the phases defined in (\ref{eq:asymp-h}), $\th_k^j$ is given by
\begin{equation}
 \theta_k^j = \int_0^{t_j}dt\,\omega(t) = j\,\th_k,
\end{equation}
with
\begin{equation}
 \th_k = 2\sqrt2\,q^{1/4}\int_0^{\pi/2}d\varphi
    \sqrt{\k^2+2q^{1/2}\sin^2\varphi},
\end{equation}
while $\vth_k$ is evaluated by use of the potential (\ref{eq:approx-potential}) as
\begin{equation}
 \vth_k = 
 2\e\,\log{\e\over{q^{1/4}\k}}-4q^{1/2}\log{{q^{-1/4}(\e+2q^{1/2})}\over\k}.
                                          \label{eq:def-vth} 
\end{equation}
With these quantities, one can evaluate the density of the particle number
in the absence of the perturbation. The result reproduces the same as
that in \cite{KLS2}, for the case of $q^{1/2}=1/(4\xi^2)\gg 1$.\par
As for the effective potential, we adopt
\begin{eqnarray}
 &&V_{11}(t)=V_{22}(t)= 2\lambda_1\Phi^2(\tanh^2\xi\tau-1), \nonumber\\
 &&V_{12}(t) = 2\lambda_2\Phi^2\,e^{i\theta(\tau)}(\tanh^2\xi\tau-1),
\end{eqnarray}
which have the same $t$-dependence as the potential in (\ref{eq:wave-eq-fk})
and vanish for large $\absv{\t}$. We will specify the $CP$-violating phase
$\th(\t)$ later when we execute numerical calculation.
If we denote
\begin{equation}
 l_1={{\lambda_1}\over{g^2}},\qquad
 l_2={{\lambda_2}\over{g^2}},
\end{equation}
the quantities composing the elements of the matrices (\ref{eq:2-dim-K-N}) are
given by
\begin{eqnarray}
 N_1 &=& -{{2l_1 q^{1/2}}\over\sqrt{1+\k^2/(4q^{1/2})}}
 \int_{-\infty}^\infty dx\,{1\over{\cosh^2 x}}
 \absv{F\left(a,b,c;{1\over{1+e^{-2x}}}\right)}^2,    \nonumber\\
 N_2 &=& -{{2l_2 q^{1/2}}\over\sqrt{1+\k^2/(4q^{1/2})}}
 \int_{-\infty}^\infty dx\,{{\cos\th(x)}\over{\cosh^2 x}}
 \absv{F\left(a,b,c;{1\over{1+e^{-2x}}}\right)}^2,    \nonumber\\
 \e_N &=& -{1\over{N_2}}{{2l_2 q^{1/2}}\over\sqrt{1+\k^2/(4q^{1/2})}}
 \int_{-\infty}^\infty dx\,{{\sin\th(x)}\over{\cosh^2 x}}
 \absv{F\left(a,b,c;{1\over{1+e^{-2x}}}\right)}^2,    \nonumber\\
 K_1 &=& -{{2l_1 q^{1/2}}\over\sqrt{1+\k^2/(4q^{1/2})}}
 \int_{-\infty}^\infty dx\,{1\over{\cosh^2 x}}\left(e^x+e^{-x}\right)^{4i\e}
 F\left(a,b,c;{1\over{1+e^{-2x}}}\right)^2,   \nonumber\\
 K_2 &=& -{{2l_2 q^{1/2}}\over\sqrt{1+\k^2/(4q^{1/2})}}
 \int_{-\infty}^\infty dx\,{{\cos\th(x)}\over{\cosh^2 x}}
 \left(e^x+e^{-x}\right)^{4i\e}
 F\left(a,b,c;{1\over{1+e^{-2x}}}\right)^2,   \nonumber\\
 \e_K &=& -{1\over{K_2}}{{2l_2 q^{1/2}}\over\sqrt{1+\k^2/(4q^{1/2})}}
 \int_{-\infty}^\infty dx\,{{\sin\th(x)}\over{\cosh^2 x}}
 \left(e^x+e^{-x}\right)^{4i\e}
 F\left(a,b,c;{1\over{1+e^{-2x}}}\right)^2. \nonumber\\
\end{eqnarray}
\par
Now we are ready to evaluate the densities of the particle number and charges.
The remaining quantities we must specify are the values of $q$, $l_1$ and $l_2$,
and the functional form of $\th(\t)$. All the quantities with mass dimensions
are expressed with appropriate powers of the inflaton mass $m$.
For definiteness, we adopt
\begin{equation}
 \theta(x) = {{\theta_0}\over{1+e^{2x}}},
\end{equation}
which changes from $\theta_0$ to $0$ as $x$ goes from $-\infty$ to
$\infty$. We take $q=200$, for which the parametric resonance predicted by
the Mathieu function occurs for a wide range of the momentum $\k$, that is,
in the broad resonance. We numerically evaluate the scattering matrices
(\ref{eq:B-matrix-for-zero-W}) for a fixed $\k$, and study the time evolution
of
\begin{eqnarray}
 n_k^j &=& 
 \sum_{a,b=1}^2\left(\absv{\b_{ab}^j}^2+\absv{\tb_{ab}^j}^2\right),\nonumber\\
 j_{1k}^j &=& 
 \absv{\tb_{11}^j}^2+\absv{\tb_{21}^j}^2-\absv{\b_{11}^j}^2-\absv{\b_{21}^j}^2,
                     \label{eq:def-j1-k}\\
 j_{2k}^j &=& 
 \absv{\tb_{12}^j}^2+\absv{\tb_{22}^j}^2-\absv{\b_{12}^j}^2-\absv{\b_{22}^j}^2,
                     \label{eq:def-j2-k}
\end{eqnarray}
where (\ref{eq:def-j1-k}) and (\ref{eq:def-j2-k}) are the charge densities defined
with $Q_1=\pmatrix{1&0\cr 0&0}$ and $Q_2=\pmatrix{0&0\cr 0&1}$ in (\ref{eq:j-k-j}),
respectively, by numerically solving the descent equations (\ref{eq:ab-j+1-by-j})
with the initial condition $\a^0=\ta^0=1$ and $\b^0=\tb^0=0$.
Because of the $U(1)$ symmetry mentioned above (\ref{eq:B-matrix-for-zero-W}),
$j_{1k}^j+j_{2k}^j=0$ in any $j$-th interval.
Although the particle number grows exponentially for $\k$ in one of the resonance
band, it ceases to grow due to the back reaction after a finite number of inflaton
oscillations, as studied in \cite{KLS2}.
How many times the inflaton oscillates before the particle creation ends
depends on $q$ and the coupling constant $g$.
For simplicity, we use the densities in the tenth interval when we evaluate the 
total densities of the generated particles and charges, since 10 is a typical
number of the oscillations before the preheating ends.\par
\begin{figure}[ht]
 \epsfysize=6cm
 \centerline{\epsfbox{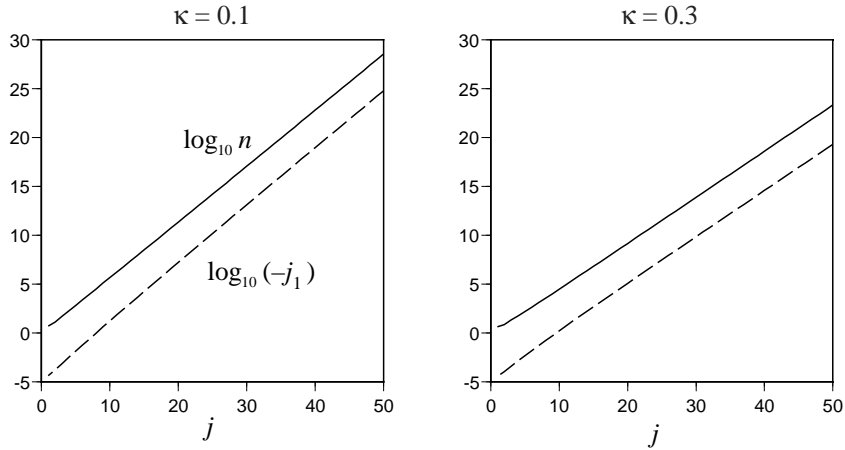}}
 \caption{Time evolutions of the densities $n_k$ (solid curve) and 
 $j_{1k}$ (dashed curve) for $\k=0.1$ and $0.3$.}
 \label{fig:densities-j-1}
\end{figure}
\begin{figure}[ht]
 \epsfysize=6cm
 \centerline{\epsfbox{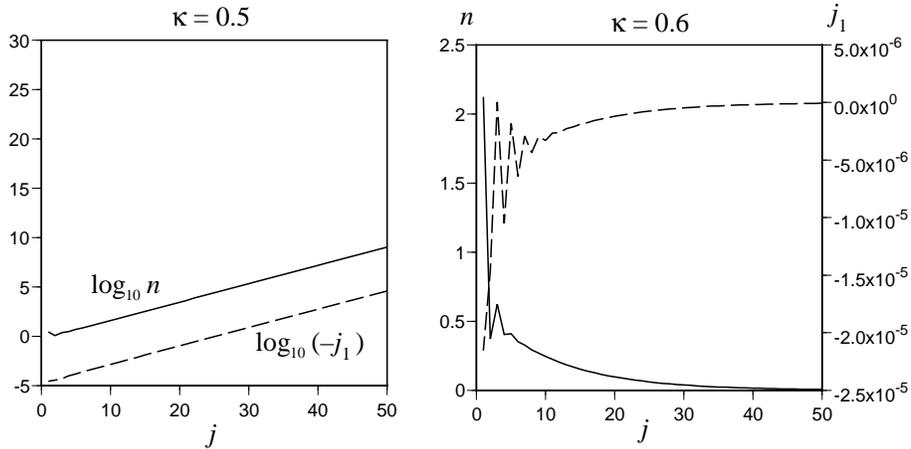}}
 \caption{Time evolutions of the densities $n_k$ (solid curve)  and
 $j_{1k}$ (dashed curve) for $\k=0.5$ and $0.6$.}
 \label{fig:densities-j-2}
\end{figure}

Now let us show the numerical results.
At first, neglecting the back reaction, we studied the time evolution of the densities
for $\theta_0=10^{-3}$, $l_1=0.01$ and $l_2=0.02$.
Since $\b_k$ in (\ref{eq:unperturb-a-b}) behaves as $\absv{\b_k}\simeq e^{-\pi\k^2}$,
the modes with $\k$ satisfying $\k\ltsim 1/\sqrt\pi\simeq 0.56$ are expected to 
belong to the first resonance band.
The time evolution of $n_k$ and $j_{1k}$ for several $\k$'s are shown in
Figs.~\ref{fig:densities-j-1} and \ref{fig:densities-j-2}.
As expected, the densities for $\k\ltsim 0.5$ grow exponentially, and
$\absv{j_{1k}^j}\sim 10^{-4}\, n_k^j$ for this choice of the parameters.\par
\begin{figure}[ht]
 \epsfysize=8.0cm
 \centerline{\epsfbox{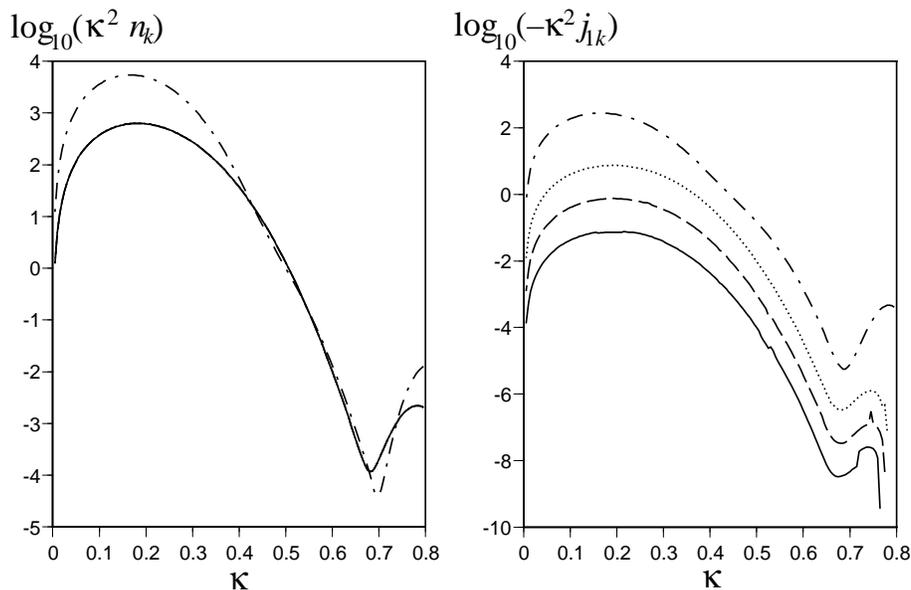}}
 \caption{The densities of the particle number $\k^2 n_k$ and charge 
 $-\k^2 j_{1k}$ at $j=10$ for $\theta_0=10^{-3}$ (solid curve),
 $10^{-2}$ (dashed curve), $10^{-1}$ (dotted curve) and $\pi$
 (dashed-dotted curve).}
 \label{fig:theta0-dep}
\end{figure}

We estimate the total amount of densities of the generated particles and charges
by integrating $n_k^j$ and $j_{1k}^j$ at $j=10$ over $k$:
\begin{eqnarray}
 n&=& \int d^3\bm{k}\,n_k = 
 8\sqrt2 \pi\,m^3\,q^{3/4}\int_0^\infty d\k\,\k^2\,n_k,   \nonumber\\
 j_1 &=&
 8\sqrt2 \pi\,m^3\,q^{3/4}\int_0^\infty d\k\,\k^2\,j_{1k},   \nonumber\\
 j_2 &=&
 8\sqrt2 \pi\,m^3\,q^{3/4}\int_0^\infty d\k\,\k^2\,j_{2k}.
                   \label{eq:total-number-charge}
\end{eqnarray}
We show the behavior of the integrands at $j=10$ as functions of $\k$ for 
several values of $\theta_0$ in Fig.~\ref{fig:theta0-dep}.
Except for the case of $\theta_0=\pi$, the generated particle numbers are not
affected by the $CP$ violation, while the generated charge is almost proportional
to $\theta_0$.
As seen from the $\k$-dependence of the integrands of (\ref{eq:total-number-charge}) 
in Fig.~\ref{fig:theta0-dep}, the second resonance band exists near $\k=0.8$ but
its contribution is about five orders smaller than that from the first resonance band.
We performed the calculation up to $\k=10$ and found several higher resonance bands,
whose contributions are negligible.
The values of the integrals in (\ref{eq:total-number-charge}) are listed
in Table~\ref{tab:integrals-theta0}.\par
\begin{table}
\begin{center}
 \begin{tabular}{c|cc}
 \hline
  $\theta_0$ & $\int d\k\,\k^2 n_k$  & $\int d\k\,\k^2 j_{1k}$  \\
 \hline
  $10^{-3}$  & $130.5096$   &  $-1.609334\times10^{-2}$  \\
  $10^{-2}$  & $130.5156$   &  $-1.544579\times10^{-1}$  \\
  $10^{-1}$  & $131.1163$   &  $-1.537716$   \\
  $\pi$      & $990.7411$   &  $-50.84228$   \\
 \hline 
 \end{tabular}
 \caption{The integrals appearing in the total densities of particle number
 and charges (\ref{eq:total-number-charge}) for several values of $\theta_0$.}
 \label{tab:integrals-theta0}
\end{center}
\end{table}
We also calculated the total densities for various values of the coupling
constants $l_1$ and $l_2$ for $\theta_0=\pi$.
As shown in Table~\ref{tab:integrals-l1}, there is no regularities
in how the total amount of densities depend on the coupling constants.
\begin{table}
\begin{center}
 \begin{tabular}{c|cc}
 \hline
  $l_1$ & $\int d\k\,\k^2 n_k$  & $\int d\k\,\k^2 j_{1k}$  \\
 \hline
  $10^{-4}$ & $1.071055\times10^5$ & $2.456463\times10^2$  \\
  $10^{-3}$ & $6.773442\times10^4$ & $-2.316711\times10^3$  \\
  $10^{-2}$ & $9.907411\times10^2$ & $-5.084228\times10^1$   \\
  $10^{-1}$ & $1.656527\times10^3$ & $-2.069592\times10^2$   \\
 \hline 
 \end{tabular}
 \caption{The integrals appearing in the total densities of particle number
  and charges (\ref{eq:total-number-charge}) for several values of $l_1$ and 
  $l_2=2l_1$ with $\theta_0=\pi$.}
 \label{tab:integrals-l1}
\end{center}
\end{table}
\section{Discussions}
Since all the particles at present are thought to be created after
the reheating process in the inflationary scenario and the process is
far from equilibrium, it is natural to consider that the asymmetry between
matter and antimatter was also generated in that era.
We formulated charge generation during the preheating, which is viewed as
a particle creation in the oscillating background.\par
If the scale factor of the FRW universe changes from $R_{\rm preheating}$
to $R_{\rm th}$ just after thermalization, the 
charge-to-entropy ratio will be given by
\begin{equation}
 {{j_1}\over s} =
 {{8\sqrt2 \pi\,m^3\,q^{3/4}\int_0^\infty d\k\,\k^2\,j_{1k}}\over
  {{{2\pi^2}\over{45}}g_{*S}T_{\rm th}^3}}
  \left({{R_{\rm preheating}}\over{R_{\rm th}}}\right)^3,
\end{equation}
where $g_{*S}$ counts the degrees of freedom contributing to the entropy
and $T_{\rm th}$ is the temperature just after thermalization.
So far, we have not specified what the charge is, but we expect it to be 
conserved after the preheating process.
One may think it as the baryon or lepton number, regarding the $\chi$-field
as the scalar partner of the quarks or leptons.
If the preheating occurs before the electroweak phase transition,
the generated lepton number is converted to the baryon number through the
sphaleron process.
When the charge is the weak hypercharge, the equilibrium state after
thermalization with the hypercharge excess will become baryon-rich state,
just as predicted in the theory of electroweak baryogenesis.
Anyway, our next subject will be to construct a realistic model, including
at least two complex scalars, which admit
the charge generation based on the mechanism studied here and to evaluate
the baryon asymmetry produced by that model.\par
Although we have considered the coherently oscillating field as the inflaton,
the formalism may be applicable to the Affleck-Dine scalar, which is coherently
rotating in the internal space with a constant frequency if we neglect the
damping due to the expansion of the universe.
The AD scalar, which is the classical squark or a slepton field, is thought
to decay perturbatively to quarks and leptons to leave $B$ or $L$. 
This situation is similar to that of the conventional reheating process and
one may expect that there is also a process similar to the preheating in
the AD mechanism.
\section*{Acknowledgments}
The authors thank J.~Sato for valuable discussions.
This work is supported in part by Grant-in-Aid for Scientific
Research on Priority Areas (Physics of $CP$ violation, No.12014212)
from the Ministry of Education, Science, and Culture of Japan.
%
%

%
%

\baselineskip=13pt
\begin{thebibliography}{99}
%
\bibitem{KLS1} L.~Kofman and A.~Linde and A.~A.~Starobinsky, \PRLet{73}{94}{3195}.
\bibitem{Shapsh} J.~Garcia-Bellido, D.~Grigoriev, A.~Kusenko and M.~Shaposhnikov,
\PRD{60}{99}{123504}.
\bibitem{KLS2} L.~Kofman and A.~Linde and A.~A.~Starobinsky,
\PRD{56}{97}{3258}.
\bibitem{LandauQM} L.~D.~Landau and E.~M.~Lifshitz, Mechanics, Section~27.
\bibitem{FKOTT} See, for example, K.~Funakubo, A.~Kakuto, S.~Otsuki, K.~Takenaga
and F.~Toyoda, \PRD{50}{94}{1105}.
%
\end{thebibliography}
\end{document}